\documentclass[twocolumn,aps,showpacs]{revtex4}
\usepackage{mathrsfs}
\usepackage{graphicx}
\begin{document}
\title {Analytical solutions to the spin-1 Bose-Einstein condensates}
\author{Zhi-Hai Zhang}
\author{Cong Zhang}
\author{Shi-Jie Yang\footnote{Corresponding author: yangshijie@tsinghua.org.cn}}
\author{Shiping Feng}
\affiliation{Department of Physics, Beijing Normal University,
Beijing 100875, PR China}
\begin{abstract}
We analytically solve the one-dimensional coupled Gross-Pitaevskii equations which govern the motion of
$F=1$ spinor Bose-Einstein condensates. The nonlinear density-density interactions are decoupled by
making use of the unique properties of the Jacobian elliptical functions. Several types of complex
stationary solutions are deduced. Furthermore, exact non-stationary solutions to the time-dependent
Gross-Pitaevskii equations are constructed by making use of the spin-rotational symmetry of the
Hamiltonian. The spin-polarizations exhibit kinked configurations. Our method is applicable to other
coupled nonlinear systems.
\end{abstract}
\pacs{03.75.Mn, 03.75.Kk, 67.85.Fg, 67.85.De} \maketitle

\section{Introduction}
Most Bose-Einstein condensates (BECs) of dilute atomic gases
realized thus far have internal degrees of freedom arising from
spins. The spinor BECs have been realized in the optical traps which
confine the atoms regardless of their spin hyperfine states. The
direction of the spin can change dynamically due to collisions
between the atoms\cite{Barrett,Gorlitz,Leanhardt}. The spinor BECs
exhibit a rich variety of magnetic phenomena. They give rise to
phenomena that are not present in the single component BECs,
including magnetic crystallization, spin textures as well as
fractional vortices\cite{Stenger,Miesner,Kobayashi}. Moreover, there
exists an interplay between superfluidity and magnetism due to the
spin-gauge symmetry. A ferromagnetic BEC spontaneously creates a
supercurrent as the spin is locally rotated\cite{Ueda,Kawaguchi}.

A spinor condensate formed by atoms with spin $F=1$ is described by
a macroscopic wave function with three components
$\Psi=(\psi_{+1},\psi_0,\psi_{-1})^T$. The mean-field Hamiltonian is
expressed as\cite{Ho,Chang}
\begin{eqnarray}
H=\int d\textbf{r}&&\{\sum_{m=-1}^{1}\psi_m^*[-\frac{\hbar^2}{2M}\bigtriangledown^2+V(\textbf{r})]\psi_m\nonumber\\
&&+\frac{\bar c_0}{2}n_{tot}^2+\frac{\bar
c_2}{2}|\textbf{F}|^2\},\label{hamiltonian}
\end{eqnarray}
where the spin-polarization vector $\textbf{F}=\psi_m^*
\hat{\textbf{F}}_{mn}\psi_n$ with $\hat \textbf{F}^i$ ($i=x,y,z$)
the rotational matrices. The coupling constants $\bar
c_0=(g_0+2g_2)/3$ and $\bar c_2=(g_2-g_0)/3$, with $g_F$ relating to
the $s$-wave scattering length of the total spin-$F$ channel as $g_F
=4\pi\hbar^2 a_F/M$. $n_{tot}=|\psi_1|^2+\psi_0|^2+|\psi_{-1}|^2$ is
the particle density and $V(\textbf{r})$ is the external potential.
The full symmetry of the Hamiltonian is $SO(3)\times U(1)$. The
energies are degenerate for an arbitrary state $\Psi$ and its
globally spin-rotational states $\Psi^\prime=U\Psi$, where $U$ is
the three-dimensional rotational matrix in the spin space which is
expressed by the Euler angles as $U(\alpha,\beta,\gamma)=e^{-i\hat
F_z\alpha}e^{-i\hat F_y\beta}e^{-i\hat F_z\gamma}$. In the ground
state, the symmetry is spontaneously broken in several different
ways, leading to a number of possible
phases\cite{Ho,Chang,Murata,Imambekov}.

We are concerned with the quasi-one-dimensional (1D) $F=1$ BEC in a
uniform external potential ($V(\textbf{r})=0$). The dynamical
motions of the spinor wavefunctions are governed by
$i\partial_t\psi_m=\delta H/\delta\psi_m^*$, which are explicitly
written as the coupled nonlinear Gross-Pitaevskii equations (GPEs),
\begin{equation}
i\hbar\partial_t\psi_m=[-\frac{\hbar^2}{2M}\partial_x^2+c_0n_{tot}\\
+c_2(\Psi^\dagger \hat{F}^i\Psi)
\hat{F}_{mn}^i]\psi_n,\label{temporal}
\end{equation}
where $c_0=\bar{c}_0/2a_{\perp}^2$ and $c_2=\bar{c}_2/2a_{\perp}^2$
are the reduced coupling constants with $a_{\perp}$ the transverse
width of the quasi-1D system. The GPE was first introduced for
unrelated problems\cite{Gross,Ginzburg}. It has been proved to be
effective in describing many phenomena observed in the dilute gas
BECs\cite{Dalfovo} and other quantum systems\cite{Coen,N.Z.}.

There is a lot of theoretical works that numerically solve Eqs.(\ref{temporal}) or the corresponding
stationary equations\cite{Li,Zhang,Nistazakis}. Nevertheless, it is of great interest if analytical
solutions can be acquired. By using the so-called inverse scattering transformation method and the
function transformation method\cite{Gardner,Sedawy,Uchiyama,Ieda}, some special solutions are obtained.
The soliton solutions are obtained by the similar transformation method for systems with appropriate
time or spatial variations in the coupling constant\cite{Beitia,Theocharis,Avelar,Wang}. Various
approximations are employed to study the solitons such as bright and dark solitons in the $F=1$ spinor
BECs\cite{Carr,Bradley,LLi,Yan}.

However, exact analytical solutions to the spinor BECs, especially in non-soliton forms, are absent in
literature. Attempts in this direction are usually frustrated by the complexity of the nonlinear
couplings. The challenges are two-fold: one is the nonlinear density-density interactions which is
associated with the $c_0$ term, while the other is the spin-exchange couplings between the hyperfine
states which is associated with the $c_2$ term. In this paper, we present several types of exact
solutions by decoupling the nonlinear density-density interactions. The solutions are expressed by
combinations of the Jacobian elliptic functions. Furthermore, the exact non-stationary solutions are
naturally produced by making using of the spin-rotational symmetry of the Hamiltonian.

The paper is organized as follows. Section II describes exact
stationary solutions to the coupled GPEs. In Sec. III, we present
the associated non-stationary solutions. Section IV contains a
summary of the main results.

\section{Stationary solutions}
The most general form of the stationary equations corresponding to Eqs.(\ref{temporal}) are obtained by
substituting
\begin{equation}
\left(
          \begin{array}{c}
            \psi_1(x,t) \\
            \psi_0(x,t) \\
            \psi_{-1}(x,t) \\
          \end{array}
        \right)
\rightarrow
\left(
          \begin{array}{c}
            \psi_1(x) e^{-i(\mu+\mu_1)t} \\
            \psi_0(x) e^{-i\mu t} \\
            \psi_{-1}(x) e^{-i(\mu-\mu_1)t} \\
          \end{array}
        \right),
\end{equation}
which give rise to
\begin{eqnarray}
(\mu+\mu_1)\psi_1&=&[-\frac{1}{2}\partial_x^2+(c_0+c_2)(|\psi_1|^2+|\psi_0|^2)\nonumber\\
&&+(c_0-c_2)|\psi_{-1}|^2]\psi_1+c_2\psi_0^2\psi_{-1}^*,\nonumber\\
\mu\psi_0&=&[-\frac{1}{2}\partial_x^2+(c_0+c_2)(|\psi_1|^2+|\psi_{-1}|^2)\nonumber\\
&&+c_0|\psi_0|^2]\psi_0+2c_2\psi_0^*\psi_1\psi_{-1},\nonumber\\
(\mu-\mu_1)\psi_{-1}&=&[-\frac{1}{2}\partial_x^2+(c_0+c_2)(|\psi_1|^2+|\psi_0|^2)\nonumber\\
&&+(c_0-c_2)|\psi_1|^2]\psi_{-1}+c_2\psi_0^2\psi_1^*.
\label{stationary}
\end{eqnarray}
Here we have chosen $\hbar=M=1$ as the units for convenience. The
coupling constants $c_0$ and $c_2$, or $a_0$ and $a_2$ are treated
as free parameters. It is notable that the non-zero parameter
$\mu_1$ plays an important role in the seeking of analytical
solutions.

The periodic boundary conditions
\begin{equation}
\psi_m(1)=\psi_m(0),
\hspace{4mm}\psi_m^\prime(1)=\psi_m^\prime(0),\label{boundary}
\end{equation}
are adopted in our calculation. In the following, we present four
types of complex solutions to the Eqs.(\ref{stationary}). All
solutions have been numerically verified to be stationary by the
dynamical evolution Eqs.(\ref{temporal}).

\subsection*{Type A. Solution of the \textrm{cn}-\textrm{sn} form}
Here, we focus on solutions with $\psi_0=0$. At the same time, the nonlinear density-density
interactions between different components are decoupled by using the unique properties of the Jacobian
elliptical functions. We first consider the following form of solution,
\begin{equation}
\left(
          \begin{array}{c}
             \psi_1 \\
            \psi_0 \\
            \psi_{-1} \\
          \end{array}
        \right)
=\left(
          \begin{array}{c}
            f(x)e^{i\theta(x)} \\
            0 \\
            D\textrm{sn}(kx,m) \\
          \end{array}
        \right),\label{Type A}
\end{equation}
where $f(x)=\sqrt{A+B\textrm{cn}^2(kx,m)}$. $A$, $B$ and $D$ are real constants. $\textrm{sn}$ and
$\textrm{cn}$ are the Jacobian elliptical functions and $k=4jK(m)$ with $m$ the modulus $(0<m<1)$. In
this paper we always take the number of periods $j=2$ as examples. In order to decouple the nonlinear
equations (\ref{stationary}), we make use of the property $\textrm{sn}^2+\textrm{cn}^2=1$ to have
\begin{equation}
|\psi_{-1}|^2=(1-\frac{|\psi_1|^2-A}{B})D^2,\ \
|\psi_1|^2=A+B(1-\frac{|\psi_{-1}|^2}{D^2}).
\end{equation}

Substituting the relations into the stationary Eqs.(\ref{stationary}), we obtain the decoupled equations
\begin{equation}
\tilde{\mu}_m\psi_m=-\frac{1}{2}\psi_m^{\prime\prime}+\tilde{\gamma}_m|\psi_m|^2\psi_m,
\ \ (m=0,\pm1)\label{decouple}
\end{equation}
where the effective chemical potentials $\tilde{\mu}_m$ and
interacting constants $\tilde{\gamma}_m$ are defined as
\begin{equation}
\left\{\begin{array}{ll}\tilde{\mu}_1=\mu+\mu_1-(c_0-c_2)(A+B)\frac{D^2}{B},\\
\tilde{\gamma}_1=(c_0+c_2)-(c_0-c_2)\frac{D^2}{B},
\end{array}\right.
\end{equation}
and
\begin{equation}
\left\{\begin{array}{ll}\tilde{\mu}_{-1}=\mu-\mu_1-(c_0-c_2)(A+B),\\
\tilde{\gamma}_{-1}=(c_0+c_2)-(c_0-c_2)\frac{B}{D^2}.
\end{array}\right.
\end{equation}

The Eqs.(\ref{decouple}) can be self-consistently solved as
\begin{eqnarray}
&&B=-\frac{m^2k^2}{\tilde{\gamma}_1},\nonumber\\
&&A=\frac{4\mu_1+3m^2k^2+4(c_0-c_2)B}{2(c_0+2c_2)},\nonumber\\
&&D^2=\frac{m^2k^2}{\tilde{\gamma}_{-1}},
\end{eqnarray}
with
\begin{eqnarray}
&&\mu=\frac{6c_0A+k^2(2-m^2)}{4},\nonumber\\
&&\tilde{\gamma}_1=\tilde{\gamma}_{-1}=2c_0.
\end{eqnarray}
It implies that the effective nonlinear interactions in each component should be repulsive
($\tilde\gamma_m>0$), which impose a constraint on the value of the coupling constants $c_0$.

The phase $\theta(x)$ in Eqs.(\ref{Type A}) is given by
\begin{equation}
\theta(x)=\int_0^x\frac{\alpha_1}{f(\xi)^2}d\xi,
\end{equation}
where
$\alpha_1=\pm(2\tilde{\mu}_1A^2-2\tilde{\gamma}_1A^3+k^2(1-m^2)AB)^{\frac{1}{2}}$
is an integral constant. The periodic boundary conditions
(\ref{boundary}) require that the amplitudes and phase respectively
satisfy
\begin{equation}
f(1)=f(0),\hspace{4mm} \theta(1)-\theta(0)=2j\pi\times n,
\end{equation}
where $n$ is an integer. The periodic condition for the phase $\theta(x)$ can be satisfied by properly
adjusting the modulus $m$ of the Jacobian elliptic functions.

\begin{figure}
\begin{center}
\includegraphics*[width=8.5cm]{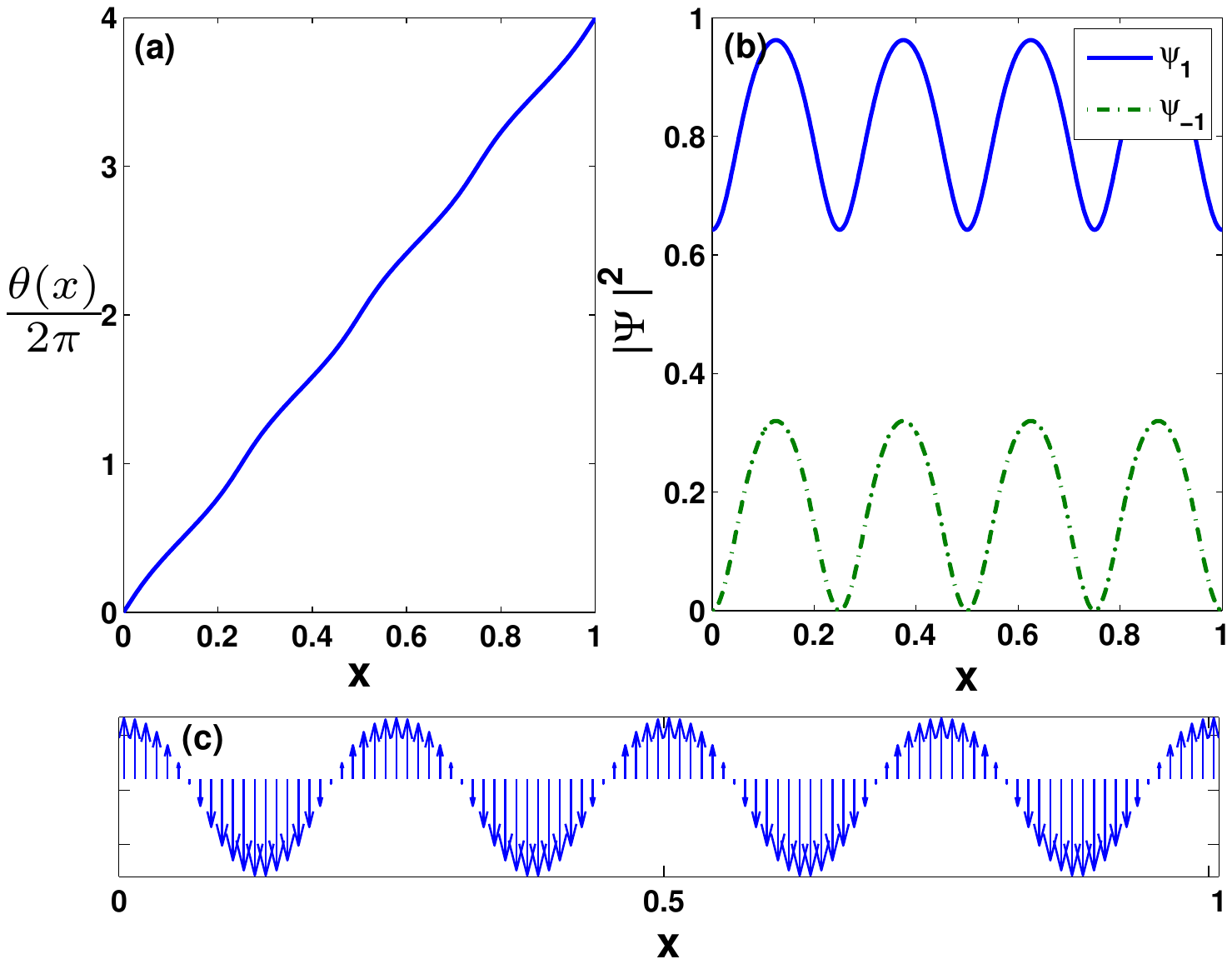}
\caption{The phase (a) and the density (b) distributions for
solution (\ref{Type A}) with $n=2$, $m=0.78$. The physical
parameters are $c_0=4$, $c_2=2$, $\mu=422.4701$, and $\mu_1=150$.
(c) The distribution of the spin-polarization.}
\end{center}
\end{figure}

Figure 1 display the distributions of some relevant physical
quantities for the solution (\ref{Type A}). We have chosen $c_0=4$,
$c_2=2$, $\mu=422.4701$ and $\mu_1=150$. Fig.1(a) is the phase
distribution with $j=2$, $n=2$ and $m=0.78$. Fig.1(b) is the density
distributions of each hyperfine state. The spin-polarization is
shown in Fig.1(c) which exhibits a kinked configuration.

\subsection*{Type B. Solution of the \textrm{sn}-\textrm{cn} form}
We take the following form of solutions to the stationary equations
(\ref{stationary}).
\begin{equation}
\left(
          \begin{array}{c}
            \psi_1 \\
            \psi_0 \\
            \psi_{-1} \\
          \end{array}
        \right)
=\left(
          \begin{array}{c}
            f(x)e^{i\theta(x)} \\
            0 \\
            D\textrm{cn}(kx,m) \\
          \end{array}
        \right),\label{Type B}
\end{equation}
where $f(x)=\sqrt{A+B\textrm{sn}^2(kx,m)}$. $A$, $B$ and $D$ are real constants. One has
\begin{equation}
|\psi_{-1}|^2=(A+B-|\psi_1|^2)\frac{D^2}{B},\ \
|\psi_1|^2=A+B-\frac{B}{D^2}|\psi_{-1}|^2.
\end{equation}
By the same means in the previous subsection, we decouple the
Eqs.(\ref{stationary}) and derive the effective chemical potentials
$\tilde{\mu}_m$ and interacting constants $\tilde{\gamma}_m$ as
\begin{equation}
\left\{\begin{array}{ll}\tilde{\mu}_1=\mu+\mu_1-(c_0-c_2)(A+B)\frac{D^2}{B},\\
\tilde{\gamma}_1=(c_0+c_2)-(c_0-c_2)\frac{D^2}{B},
\end{array}\right.
\end{equation}
and
\begin{equation}
\left\{\begin{array}{ll}\tilde{\mu}_{-1}=\mu-\mu_1-(c_0-c_2)(A+B),\\
\tilde{\gamma}_{-1}=(c_0+c_2)-(c_0-c_2)\frac{B}{D^2}.
\end{array}\right.
\end{equation}

With this form of solutions, the decoupled Eqs.(\ref{decouple}) are
self-consistently solved as
\begin{eqnarray}
&&B=\frac{m^2k^2}{\tilde{\gamma}_1},\nonumber\\
&&A=\frac{4\mu_1-3m^2k^2+4(c_0-c_2)B}{2(c_0+2c_2)},\nonumber\\
&&D^2=-\frac{m^2k^2}{\tilde{\gamma}_{-1}},\nonumber\\
\end{eqnarray}
with
\begin{eqnarray}
&&\mu=\frac{6c_0A+k^2(2-m^2)}{4},\nonumber\\
&&\tilde{\gamma}_1=\tilde{\gamma}_{-1}=2c_0.
\end{eqnarray}
One observes that $\tilde{\gamma}_m<0$ which imply the effective
interactions in each component should be attractive. The phase is
\begin{equation}
\theta(x)=\int_0^x\frac{\alpha_1}{f(\xi)^2}d\xi,
\end{equation}
where
$\alpha_1=\pm(2\tilde{\mu}_1A^2-2\tilde{\gamma}_1A^3+k^2AB)^{\frac{1}{2}}$
is an integral constant. The amplitudes and the phase should satisfy
the periodic boundary conditions (\ref{boundary}), which can be
realized by scanning the modulus $m$.

\begin{figure}
\begin{center}
\includegraphics*[width=8.5cm]{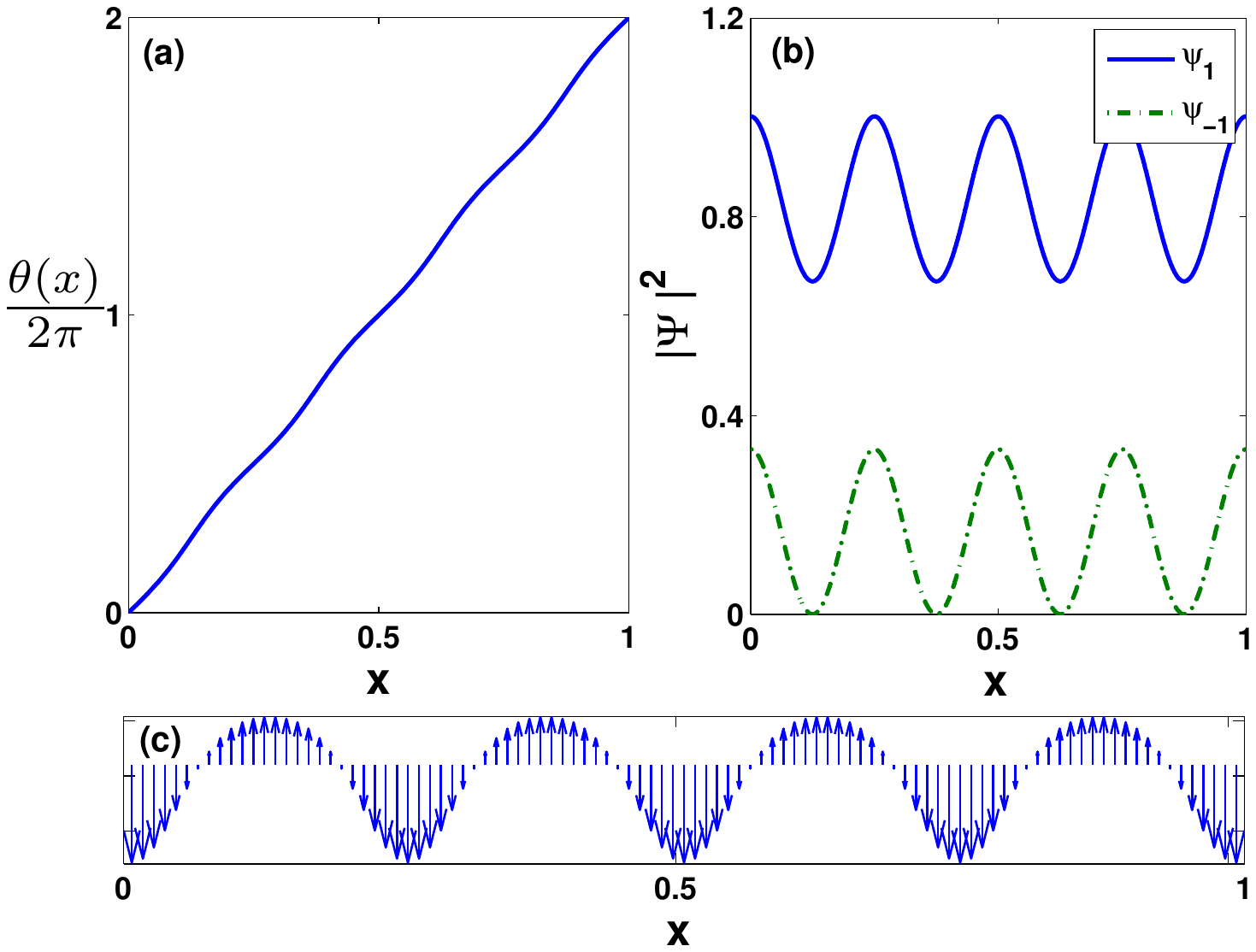}
\caption{The same as in Fig.3 for the solution (\ref{Type B}) with $n=1$ and $m=0.29$. The physical
parameters are $c_0=-2$, $c_2=6$, $\mu=47.6088$, and $\mu_1=35$.}
\end{center}
\end{figure}

Figure 2 display the distributions of the relevant physical
quantities for the solution (\ref{Type B}). We have chosen $c_0=-2$,
$c_2=6$, $\mu=47.6088$ and $\mu_1=35$. $n=1$ and the value $m=0.29$
is numerically scanned.

\subsection*{Type C. Solution of the \textrm{sn}-\textrm{dn} form}
we take the following ansatz as the most solution of the nonlinear Eqs.(\ref{stationary}).
\begin{equation}
\left(
          \begin{array}{c}
            \psi_1 \\
            \psi_0 \\
            \psi_{-1} \\
          \end{array}
        \right)
=\left(
          \begin{array}{c}
            f(x)e^{i\theta(x)} \\
            0 \\
            D\textrm{dn}(kx,m) \\
          \end{array}
        \right),\label{Type C}
\end{equation}
where $f(x)=\sqrt{A+B\textrm{sn}^2(kx,m)}$. $A$, $B$ and $D$ are real constants. One has
\begin{equation}
|\psi_{-1}|^2=\frac{D^2}{B}(B+m^2A-m^2|\psi_1|^2),\ \
|\psi_1|^2=A+\frac{B}{m^2}-\frac{B}{m^2D^2}|\psi_{-1}|^2.
\end{equation}

Similarly, the decoupled equations (\ref{decouple}) contain the
effective chemical potentials $\tilde{\mu}_m$ and interacting
constants $\tilde{\gamma}_m$ as
\begin{equation}
\left\{\begin{array}{ll}\tilde{\mu}_1=\mu+\mu_1-(c_0-c_2)(m^2A+B)\frac{D^2}{B},\\
\tilde{\gamma}_1=(c_0+c_2)-(c_0-c_2)\frac{m^2D^2}{B},
\end{array}\right.
\end{equation}
and
\begin{equation}
\left\{\begin{array}{ll}\tilde{\mu}_{-1}=\mu-\mu_1-(c_0-c_2)(A+\frac{B}{m^2}),\\
\tilde{\gamma}_{-1}=(c_0+c_2)-(c_0-c_2)\frac{B}{m^2D^2}.
\end{array}\right.
\end{equation}

This form of solutions to the decoupled Eqs.(\ref{decouple}) are
self-consistently solved as
\begin{eqnarray}
&&B=\frac{m^2k^2}{\tilde{\gamma}_1},\nonumber\\
&&A=\frac{4\mu_1-3k^2+4(c_0-c_2)\frac{B}{m^2}}{2(c_0+2c_2)},\nonumber\\
&&D^2=-\frac{k^2}{\tilde{\gamma}_{-1}},
\end{eqnarray}
with
\begin{eqnarray}
&&\mu=\frac{6c_0A-k^2(1-2m^2)}{4},\nonumber\\
&&\tilde{\gamma}_1=\tilde{\gamma}_{-1}=2c_0.
\end{eqnarray}
The negative values of $\tilde{\gamma}_m<0$ for this form of complex
solution imply effective attractive interactions in each component.
The phase is
\begin{equation}
\theta(x)=\int_0^x\frac{\alpha_1}{f(\xi)^2}d\xi,
\end{equation}
where
$\alpha_1=\pm(2\tilde{\mu}_1A^2-2\tilde{\gamma}_1A^3+k^2AB)^{\frac{1}{2}}$
is an integral constant. The periodic boundary conditions are
satisfied by properly adjusting the modulus $m$.

\begin{figure}
\begin{center}
\includegraphics*[width=8.5cm]{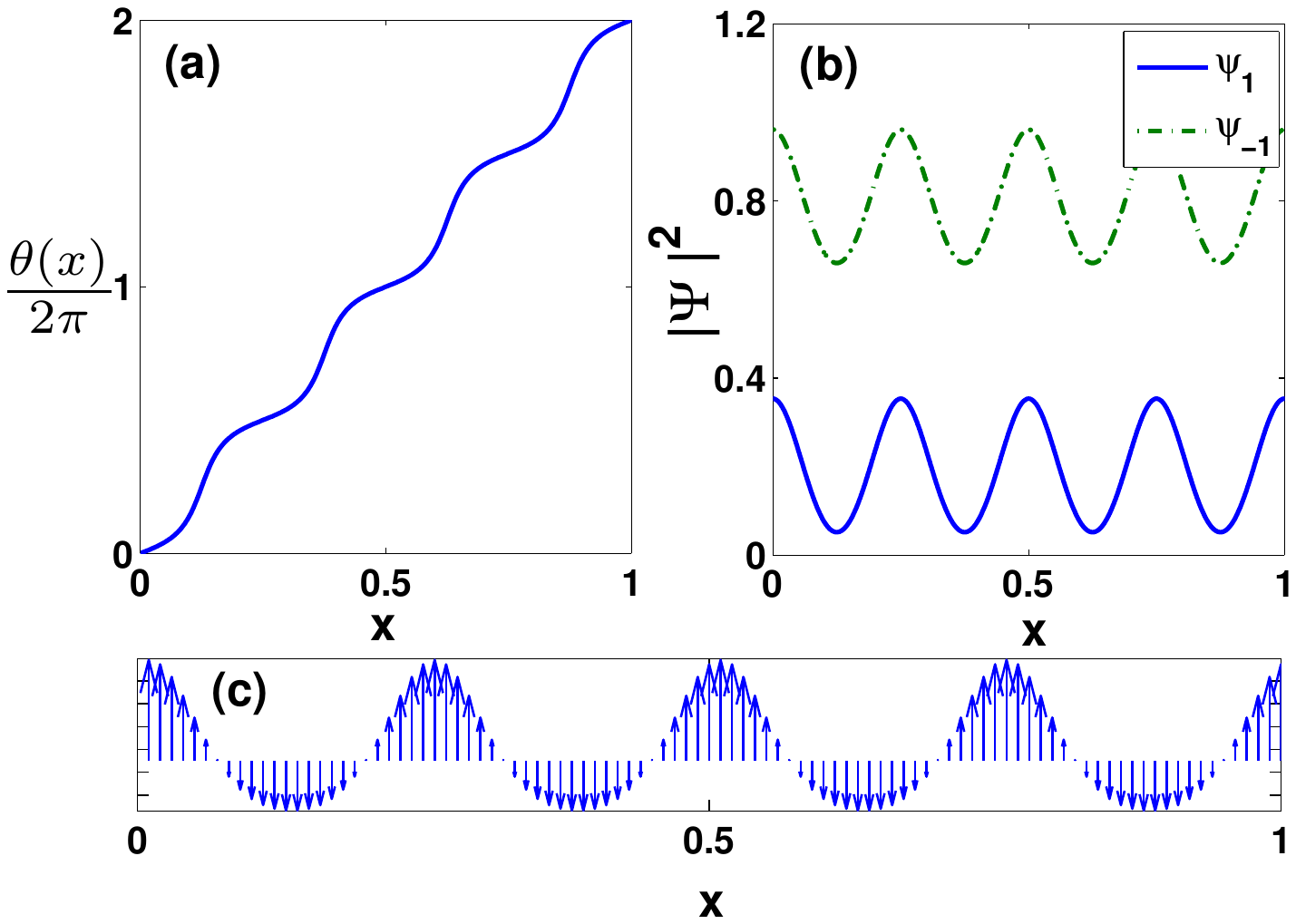}
\caption{The phase (a) and the density (b) distributions for solution (\ref{Type C}) with $n=1$,
$m=0.56$. The physical parameters are $c_0=-2$, $c_2=4$, $\mu=-70.008$, and $\mu_1=90$. (c) The
distribution of the spin-polarization.}
\end{center}
\end{figure}

Figure 3 display the distributions of the phase (Fig.3a), the
density (Fig.3b) for the solution (\ref{Type C}). The physical
parameters are chosen as $c_0=-2$, $c_2=4$, $\mu=-70.008$, and
$\mu_1=90$, whereas $n=1$ and the value $m=0.56$ is numerically
scanned. The spin polarization of this solution is shown in
Fig.3(c).

\subsection*{Type D. Solution of the \textrm{dn}-\textrm{sn} form}
Consider the following form of solutions to the nonlinear stationary
Eqs.(\ref{stationary}),
\begin{equation}
\left(
          \begin{array}{c}
            \psi_1 \\
            \psi_0 \\
            \psi_{-1} \\
          \end{array}
        \right)
=\left(
          \begin{array}{c}
            f(x)e^{i\theta(x)} \\
            0 \\
            D\textrm{sn}(kx,m) \\
          \end{array}
        \right),\label{Type D}
\end{equation}
where $f(x)=\sqrt{A+B\textrm{dn}^2(kx,m)}$. One has
\begin{equation}
|\psi_{-1}|^2=(A+B-|\psi_1|^2)\frac{D^2}{m^2B},\ \
|\psi_1|^2=A+B-\frac{m^2B}{D^2}|\psi_{-1}|^2.
\end{equation}

The stationary equations (\ref{stationary}) are decouple into
(\ref{decouple}), with the effective chemical potentials
$\tilde{\mu}_m$ and interacting constants $\tilde{\gamma}_m$,
\begin{equation}
\left\{\begin{array}{ll}\tilde{\mu}_1=\mu+\mu_1-(c_0-c_2)(A+B)\frac{D^2}{m^2B},\\
\tilde{\gamma}_1=(c_0+c_2)-(c_0-c_2)\frac{D^2}{m^2B},
\end{array}\right.
\end{equation}
and
\begin{equation}
\left\{\begin{array}{ll}\tilde{\mu}_{-1}=\mu-\mu_1-(c_0-c_2)(A+B),\\
\tilde{\gamma}_{-1}=(c_0+c_2)-(c_0-c_2)\frac{m^2B}{D^2}.
\end{array}\right.
\end{equation}

This form of solutions to the decoupled Eqs.(\ref{decouple}) are
self-consistently solved as
\begin{eqnarray}
&&B=-\frac{k^2}{\tilde{\gamma}_1},\nonumber\\
&&A=\frac{4\mu_1+3k^2+4(c_0-c_2)B}{2(c_0+2c_2)},\nonumber\\
&&D^2=\frac{m^2k^2}{\tilde{\gamma}_{-1}},\nonumber\\
\end{eqnarray}
with
\begin{eqnarray}
&&\mu=\frac{6c_0A+k^2(2m^2-1)}{4},\nonumber\\
&&\tilde{\gamma}_1=\tilde{\gamma}_{-1}=2c_0.
\end{eqnarray}
The positive values of $\tilde{\gamma}_m>0$ for this form of
solution imply that effective interactions in each component are
repulsive. The phase is
\begin{equation}
\theta(x)=\int_0^x\frac{\alpha_1}{f(\xi)^2}d\xi,
\end{equation}
where
$\alpha_1=\pm(2\tilde{\mu}_1A^2-2\tilde{\gamma}_1A^3+k^2(m^2-1)AB)^{\frac{1}{2}}$
is an integral constant. Accordingly, the amplitudes and phase
should satisfy the periodic boundary conditions (\ref{boundary}).

\begin{figure}
\begin{center}
\includegraphics*[width=8.5cm]{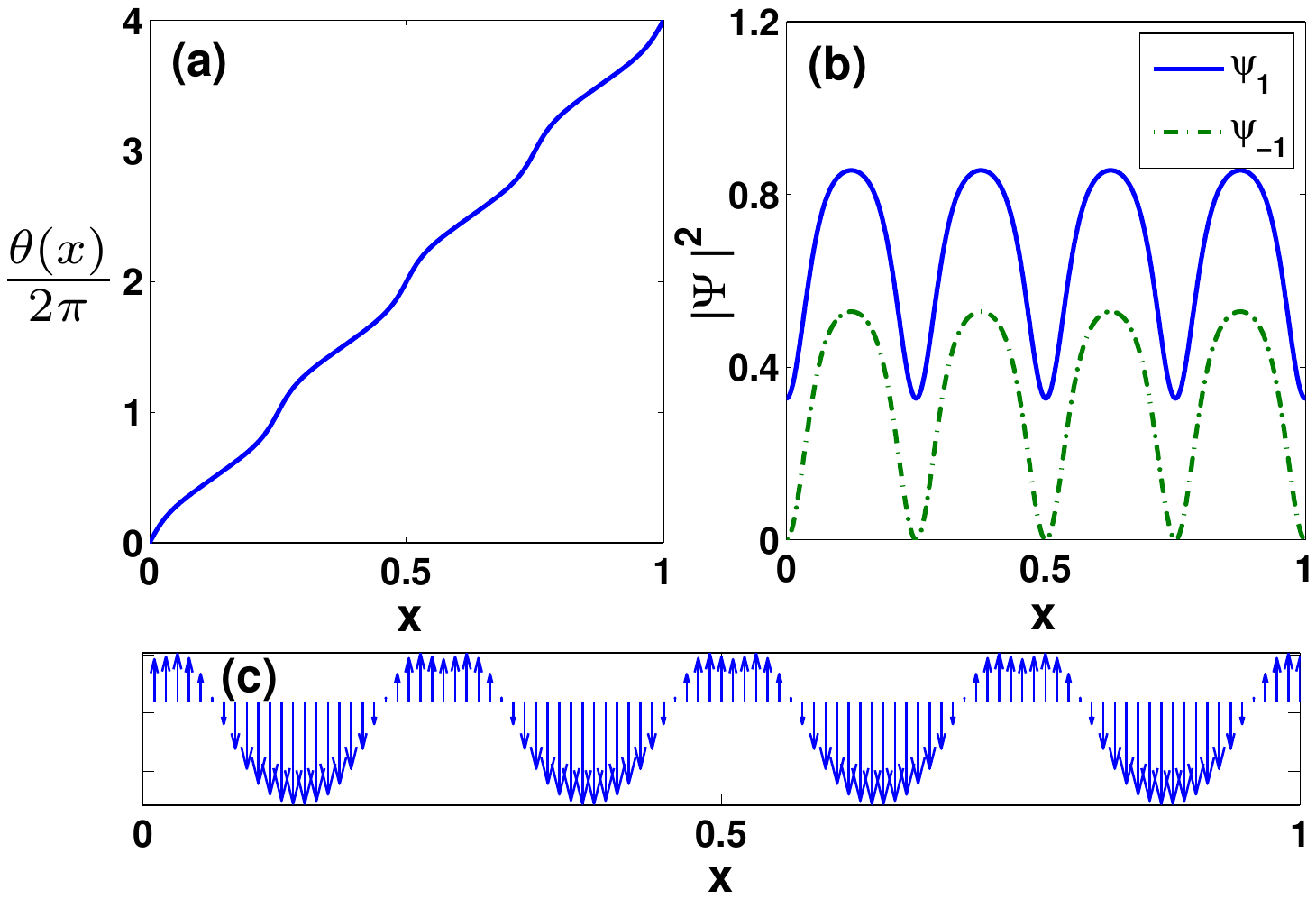}
\caption{The same as in Fig.3 for the solution (\ref{Type D}) with $n=2$ and $m=0.95$. The physical
parameters are $c_0=8$, $c_2=2$, $\mu=588.3927$, and $\mu_1=90$.}
\end{center}
\end{figure}

Figure 4 display the distributions of the phase (Fig.4a), the
density (Fig.4b) for the solution (\ref{Type D}). The physical
parameters are chosen as $c_0=8$, $c_2=2$, $\mu=588.3927$, and
$\mu_1=90$, whereas $n=2$ and the value $m=0.95$ is numerically
scanned. The spin-polarization of this solution is shown in
Fig.4(c).

\section{Non-stationary solutions}
Suppose $\Psi(x)=(\psi_1(x),\psi_0(x),\psi_{-1}(x))^T$ is a
stationary solution to the equations (\ref{stationary}), the full
spatial and temporal form of the solution is written by adding the
time factor,
\begin{equation}
\Psi(x,t)=\left(
          \begin{array}{c}
            \psi_1(x) e^{-i(\mu+\mu_1)t} \\
            \psi_0(x) e^{-i\mu t} \\
            \psi_{-1}(x) e^{-i(\mu-\mu_1)t} \\
          \end{array}
        \right),
\end{equation}
Since the Hamiltonian (\ref{hamiltonian}) has the $ SO(3)$
spin-rotational symmetry, then $\Psi^\prime(x,t)=U\Psi(x,t)$ is a
solution to the equations of motion (\ref{temporal}). Here $U$ is
the three-dimensional rotational matrix in the spin space expressed
by the three Euler angles $(\alpha,\beta,\gamma)$,
\begin{equation}
U= \left(
          \begin{array}{ccc}
            e^{-i(\alpha+\gamma)}\cos^2\frac{\beta}{2} & -\frac{1}{\sqrt{2}}e^{-i\alpha}\sin\beta & e^{-i(\alpha-\gamma)}\sin^2\frac{\beta}{2} \\
            \frac{1}{\sqrt{2}}e^{-i\gamma}\sin\beta & \cos\beta & -\frac{1}{\sqrt{2}}e^{i\gamma}\sin\beta \\
            e^{i(\alpha-\gamma)}\sin^2\frac{\beta}{2} & \frac{1}{\sqrt{2}}e^{i\alpha}\sin\beta & e^{i(\alpha+\gamma)}\cos^2\frac{\beta}{2} \\
          \end{array}
        \right).
\end{equation}
Given the non-zero value of $\mu_1$, we observe that
$\Psi^\prime(x,t)$ is an exact non-stationary solution. It is
notable that $\Psi^\prime(x,0)$ is no longer the stationary solution
to the Eqs.(\ref{stationary}). In this section, we construct the
corresponding non-stationary solutions associated to the stationary
solutions obtained in the previous section. We will fix the Euler
angles $\alpha=7\pi/13$, $\beta=\pi/4$ and $\gamma=\pi/11$.

\subsection{Solution associated to Type A}
By applying the spin-rotational transformation to the stationary
solution (\ref{Type A}), we obtain the the non-stationary solution
as follows:
\begin{widetext}
\begin{eqnarray}
\Psi^\prime(x,t)= \left(
          \begin{array}{c}
            \psi_1e^{-i(\mu+\mu_1)t}e^{-i(\alpha+\gamma)}\cos^2(\frac{\beta}{2})+\psi_{-1}e^{-i(\mu-\mu_1)t}e^{-i(\alpha-\gamma)}\sin^2(\frac{\beta}{2}) \\
            \psi_1e^{-i(\mu+\mu_1)t}e^{-i\gamma}\sin\beta/\sqrt{2}-\psi_{-1}e^{-i(\mu-\mu_1)t} e^{i\gamma}\sin\beta/\sqrt{2} \\
            \psi_1e^{-i(\mu+\mu_1)t}e^{i(\alpha-\gamma)}\sin^2(\frac{\beta}{2})+\psi_{-1}e^{-i(\mu-\mu_1)t}e^{i(\alpha+\gamma)}\cos^2(\frac{\beta}{2}) \\
          \end{array}
        \right)\label{non-A}
\end{eqnarray}
\end{widetext}
where $\psi_1=\sqrt{A+B\textrm{cn}^2(kx,m)}e^{i\theta(x)}$ and
$\psi_{-1}=D\textrm{sn}(kx,m)$. The density of each component
$n_i(x,t)=|\psi_i^\prime(x,t)|^2$, $(i=1,2,3)$ is given by

\begin{widetext}
\begin{eqnarray}
n_1(x,t)&=&|\psi_1|^2(\cos^2\frac{\beta}{2})^2+|\psi_{-1}|^2(\sin^2\frac{\beta}{2})^2+2\sqrt{A+B\textrm{cn}^2}D\textrm{sn}\sin^2\frac{\beta}{2}\cos^2\frac{\beta}{2}\cos(2\mu_1t-2\gamma+\theta),\nonumber\\
n_0(x,t)&=&|\psi_1|^2\sin^2\beta/2+|\psi_{-1}|^2\sin^2\beta/2-\sqrt{A+B\textrm{cn}^2}D\textrm{sn}\sin^2\beta \cos(2\mu_1t-2\gamma+\theta),\nonumber\\
n_{-1}(x,t)&=&|\psi_1|^2(\sin^2\frac{\beta}{2})^2+|\psi_{-1}|^2(\cos^2\frac{\beta}{2})^2+2\sqrt{A+B\textrm{cn}^2}D\textrm{sn}\sin^2\frac{\beta}{2}\cos^2\frac{\beta}{2}\cos(2\mu_1t-2\gamma+\theta).\nonumber\\
\end{eqnarray}
\end{widetext}

\begin{figure}
\begin{center}
\includegraphics*[width=8.5cm]{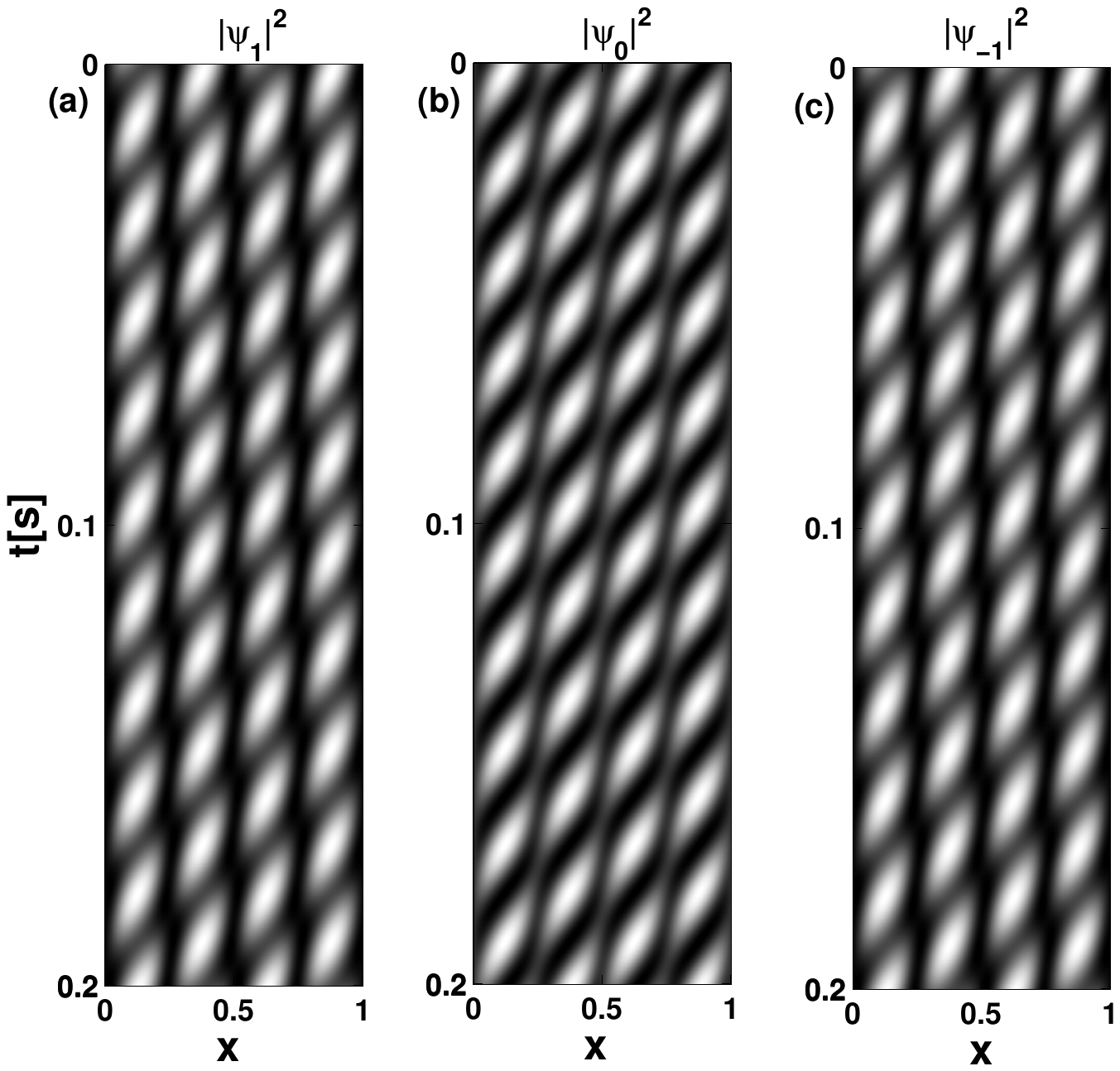}
\caption{The density distribution of the non-stationary solution (\ref{non-A}). The parameters are the
same as in Fig.1.}
\end{center}
\end{figure}

\begin{figure}
\begin{center}
\includegraphics*[width=8.5cm]{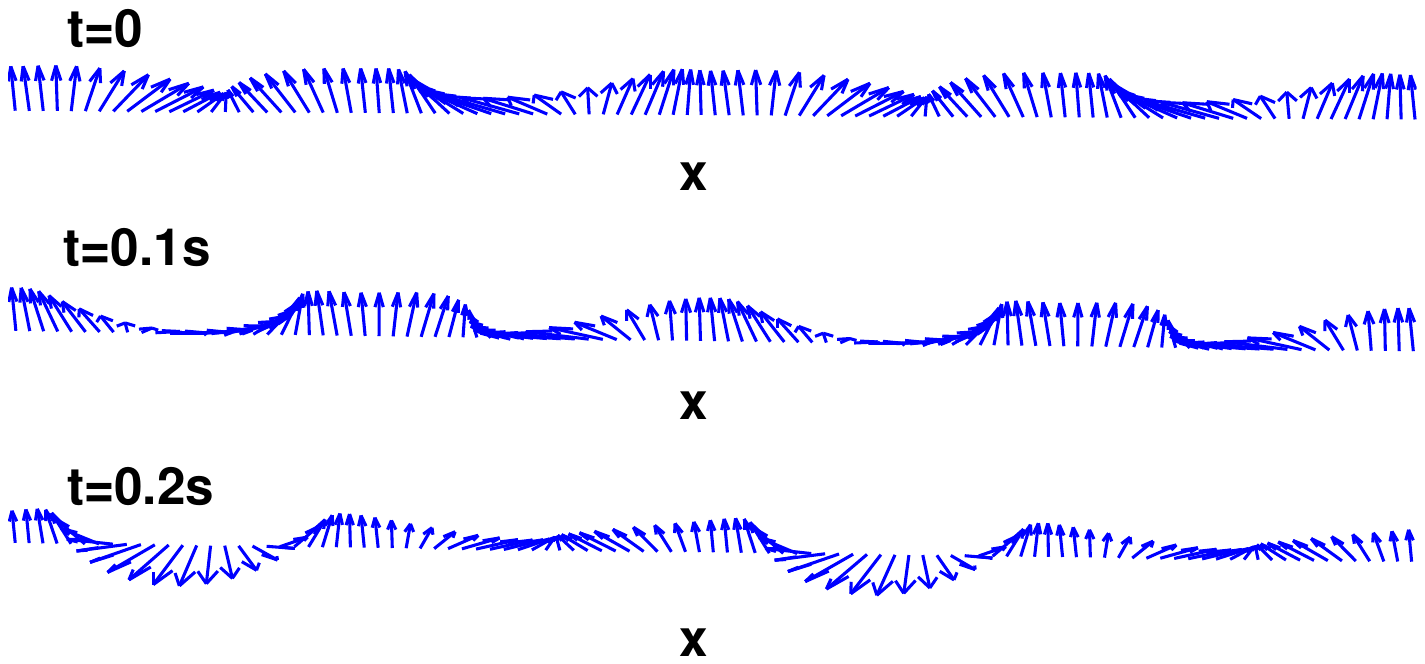}
\caption{Snapshots of the spin-polarization distributions of the non-stationary solutions (\ref{non-A})
at $t=0s, t=0.1s$ and $t=0.2s$, respectively.}
\end{center}
\end{figure}

\subsection{Solution associated to Type B}
By applying the spin-rotational transformation to the stationary
solution (\ref{Type B}), we obtain the non-stationary solution as
follows:\begin{widetext}
\begin{eqnarray}
\Psi^\prime(x,t)= \left(
          \begin{array}{c}
            \psi_1e^{-i(\mu+\mu_1)t}e^{-i(\alpha+\gamma)}\cos^2(\frac{\beta}{2})+\psi_{-1}e^{-i(\mu-\mu_1)t}e^{-i(\alpha-\gamma)}\sin^2(\frac{\beta}{2}) \\
            \psi_1e^{-i(\mu+\mu_1)t}e^{-i\gamma}\sin\beta/\sqrt{2}-\psi_{-1}e^{-i(\mu-\mu_1)t} e^{i\gamma}\sin\beta/\sqrt{2} \\
            \psi_1e^{-i(\mu+\mu_1)t}e^{i(\alpha-\gamma)}\sin^2(\frac{\beta}{2})+\psi_{-1}e^{-i(\mu-\mu_1)t}e^{i(\alpha+\gamma)}\cos^2(\frac{\beta}{2}) \\
          \end{array}
        \right),\label{non-B}
\end{eqnarray}
\end{widetext}
where $\psi_1=\sqrt{A+B\textrm{sn}^2(kx,m)}e^{i\theta(x)}$ and
$\psi_{-1}=D\textrm{cn}(kx,m)$. The density of each component
$n_i(x,t)=|\psi_i^\prime(x,t)|^2$, $(i=1,2,3)$ is given by
\begin{widetext}
\begin{eqnarray}
n_1(x,t)&=&|\psi_1|^2(\cos^2\frac{\beta}{2})^2+|\psi_{-1}|^2(\sin^2\frac{\beta}{2})^2+2\sqrt{A+B\textrm{sn}^2}D\textrm{cn}\sin^2\frac{\beta}{2}\cos^2\frac{\beta}{2}\cos(2\mu_1t-2\gamma+\theta),\nonumber\\
n_0(x,t)&=&|\psi_1|^2\sin^2\beta/2+|\psi_{-1}|^2\sin^2\beta/2-\sqrt{A+B\textrm{sn}^2}D\textrm{cn}\sin^2\beta \cos(2\mu_1t-2\gamma+\theta),\nonumber\\
n_{-1}(x,t)&=&|\psi_1|^2(\sin^2\frac{\beta}{2})^2+|\psi_{-1}|^2(\cos^2\frac{\beta}{2})^2+2\sqrt{A+B\textrm{sn}^2}D\textrm{cn}\sin^2\frac{\beta}{2}\cos^2\frac{\beta}{2}\cos(2\mu_1t-2\gamma+\theta).\nonumber\\
\end{eqnarray}
\end{widetext}

\begin{figure}
\begin{center}
\includegraphics*[width=8.5cm]{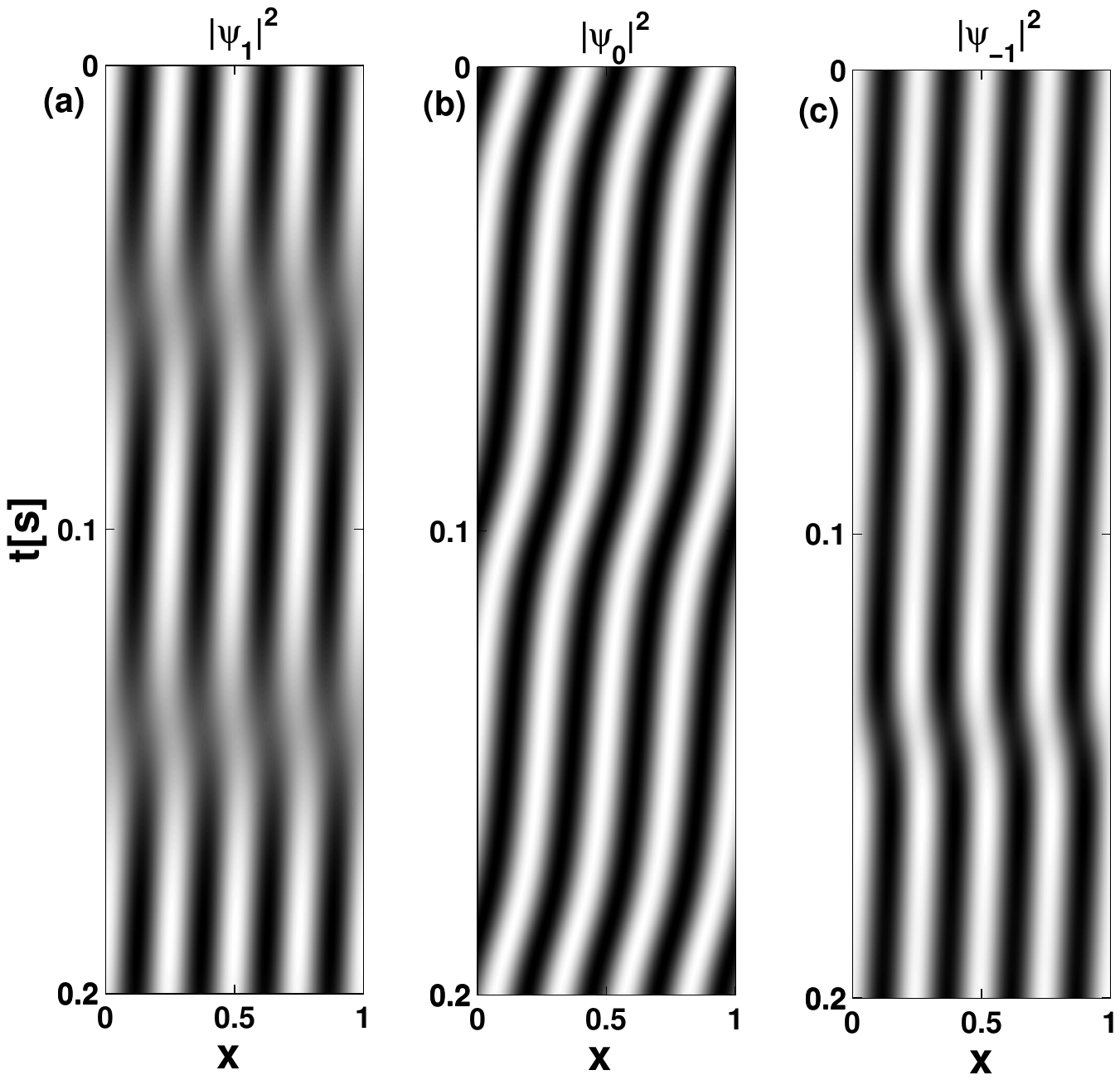}
\caption{The density distribution of the non-stationary solution (\ref{non-B}). The parameters are the
same as in Fig.2.}
\end{center}
\end{figure}

\begin{figure}
\begin{center}
\includegraphics*[width=8.5cm]{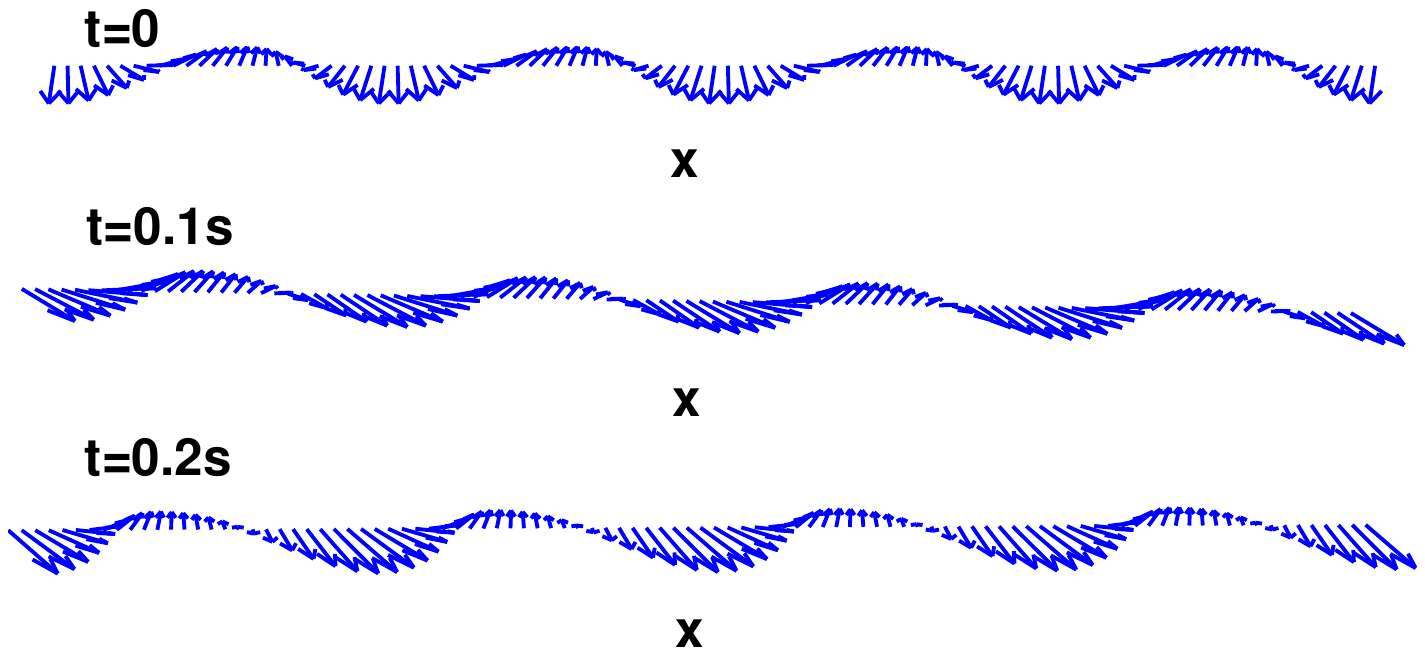}
\caption{Snapshots of the spin-polarization distributions of the non-stationary solutions (\ref{non-B})
at $t=0s, t=0.1s$ and $t=0.2s$, respectively.}
\end{center}
\end{figure}

Fig.(5) and Fig.(7) depict the temporal evolution of the density in
each component from the non-stationary solutions (\ref{non-A}) and
(\ref{non-B}). They exhibit periodical structures in both the time
and the space. Fig.(6) and Fig.(8) describe the evolution of the
spin-polarizations for the non-stationary solutions (\ref{non-A})
and (\ref{non-B}) at $t=0, t=0.1s$ and $t=0.2s$, respectively.

\subsection{Solution associated to Type C}
By applying the spin-rotational transformation to the stationary
solution (\ref{Type C}), we obtain the the non-stationary solution
as follows:\begin{widetext}
\begin{eqnarray}
\Psi^\prime(x,t)= \left(
          \begin{array}{c}
            \psi_1e^{-i(\mu+\mu_1)t}e^{-i(\alpha+\gamma)}\cos^2(\frac{\beta}{2})+\psi_{-1}e^{-i(\mu-\mu_1)t}e^{-i(\alpha-\gamma)}\sin^2(\frac{\beta}{2}) \\
            \psi_1e^{-i(\mu+\mu_1)t}e^{-i\gamma}\sin\beta/\sqrt{2}-\psi_{-1}e^{-i(\mu-\mu_1)t} e^{i\gamma}\sin\beta/\sqrt{2} \\
            \psi_1e^{-i(\mu+\mu_1)t}e^{i(\alpha-\gamma)}\sin^2(\frac{\beta}{2})+\psi_{-1}e^{-i(\mu-\mu_1)t}e^{i(\alpha+\gamma)}\cos^2(\frac{\beta}{2}) \\
          \end{array}
        \right)\label{non-C}
\end{eqnarray}
\end{widetext}
where $\psi_1=\sqrt{A+B\textrm{sn}^2(kx,m)}e^{i\theta(x)}$ and
$\psi_{-1}=D\textrm{dn}(kx,m)$. The density of each component
$n_i(x,t)=|\psi_i^\prime(x,t)|^2$, $(i=1,2,3)$ is given by

\begin{widetext}
\begin{eqnarray}
n_1(x,t)&=&|\psi_1|^2(\cos^2\frac{\beta}{2})^2+|\psi_{-1}|^2(\sin^2\frac{\beta}{2})^2+2\sqrt{A+B\textrm{sn}^2}D\textrm{dn}\sin^2\frac{\beta}{2}\cos^2\frac{\beta}{2}\cos(2\mu_1t-2\gamma+\theta),\nonumber\\
n_0(x,t)&=&|\psi_1|^2\sin^2\beta/2+|\psi_{-1}|^2\sin^2\beta/2-\sqrt{A+B\textrm{sn}^2}D\textrm{dn}\sin^2\beta \cos(2\mu_1t-2\gamma+\theta),\nonumber\\
n_{-1}(x,t)&=&|\psi_1|^2(\sin^2\frac{\beta}{2})^2+|\psi_{-1}|^2(\cos^2\frac{\beta}{2})^2+2\sqrt{A+B\textrm{sn}^2}D\textrm{dn}\sin^2\frac{\beta}{2}\cos^2\frac{\beta}{2}\cos(2\mu_1t-2\gamma+\theta).\nonumber\\
\end{eqnarray}
\end{widetext}

\begin{figure}
\begin{center}
\includegraphics*[width=8.5cm]{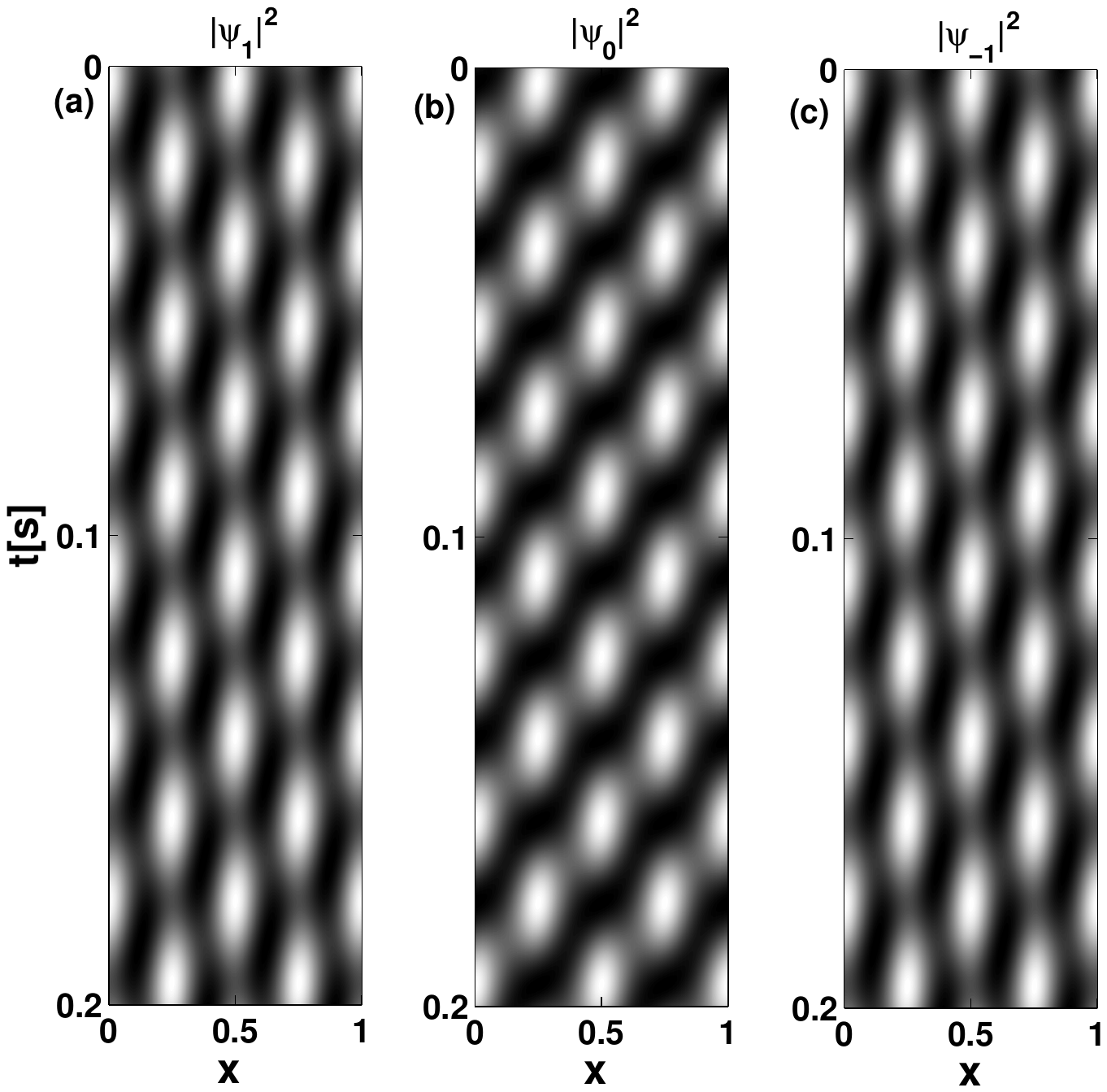}
\caption{The density distribution of the non-stationary solution (\ref{non-C}). The parameters are the
same as in Fig.3.}
\end{center}
\end{figure}

\begin{figure}
\begin{center}
\includegraphics*[width=8.5cm]{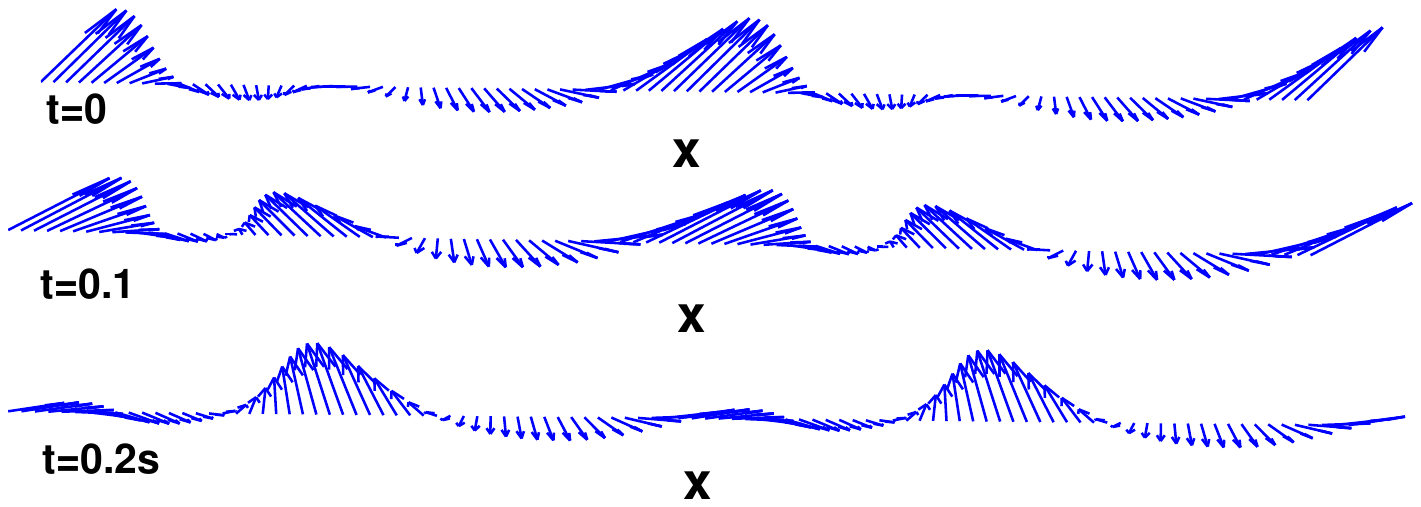}
\caption{Snapshots of the spin-polarization distributions of the non-stationary solutions (\ref{non-C})
at $t=0s, t=0.1s$ and $t=0.2s$, respectively.}
\end{center}
\end{figure}

\subsection{Solution associated to Type D}
By applying the spin-rotational transformation to the stationary
solution (\ref{Type D}), we obtain the the non-stationary solution
as follows:
\begin{widetext}
\begin{eqnarray}
\Psi^\prime(x,t)= \left(
          \begin{array}{c}
            \psi_1e^{-i(\mu+\mu_1)t}e^{-i(\alpha+\gamma)}\cos^2(\frac{\beta}{2})+\psi_{-1}e^{-i(\mu-\mu_1)t}e^{-i(\alpha-\gamma)}\sin^2(\frac{\beta}{2}) \\
            \psi_1e^{-i(\mu+\mu_1)t}e^{-i\gamma}\sin\beta/\sqrt{2}-\psi_{-1}e^{-i(\mu-\mu_1)t} e^{i\gamma}\sin\beta/\sqrt{2} \\
            \psi_1e^{-i(\mu+\mu_1)t}e^{i(\alpha-\gamma)}\sin^2(\frac{\beta}{2})+\psi_{-1}e^{-i(\mu-\mu_1)t}e^{i(\alpha+\gamma)}\cos^2(\frac{\beta}{2}) \\
          \end{array}
        \right),\label{non-D}
\end{eqnarray}
\end{widetext}
where $\psi_1=\sqrt{A+B\textrm{dn}^2(kx,m)}e^{i\theta(x)}$ and
$\psi_{-1}=D\textrm{sn}(kx,m)$. The density of each component
$n_i(x,t)=|\psi_i^\prime(x,t)|^2$, $(i=1,2,3)$ is given by
\begin{widetext}
\begin{eqnarray}
n_1(x,t)&=&|\psi_1|^2(\cos^2\frac{\beta}{2})^2+|\psi_{-1}|^2(\sin^2\frac{\beta}{2})^2+2\sqrt{A+B\textrm{dn}^2}D\textrm{sn}\sin^2\frac{\beta}{2}\cos^2\frac{\beta}{2}\cos(2\mu_1t-2\gamma+\theta),\nonumber\\
n_0(x,t)&=&|\psi_1|^2\sin^2\beta/2+|\psi_{-1}|^2\sin^2\beta/2-\sqrt{A+B\textrm{dn}^2}D\textrm{sn}\sin^2\beta \cos(2\mu_1t-2\gamma+\theta),\nonumber\\
n_{-1}(x,t)&=&|\psi_1|^2(\sin^2\frac{\beta}{2})^2+|\psi_{-1}|^2(\cos^2\frac{\beta}{2})^2+2\sqrt{A+B\textrm{dn}^2}D\textrm{sn}\sin^2\frac{\beta}{2}\cos^2\frac{\beta}{2}\cos(2\mu_1t-2\gamma+\theta).\nonumber\\
\end{eqnarray}
\end{widetext}

\begin{figure}
\begin{center}
\includegraphics*[width=8.5cm]{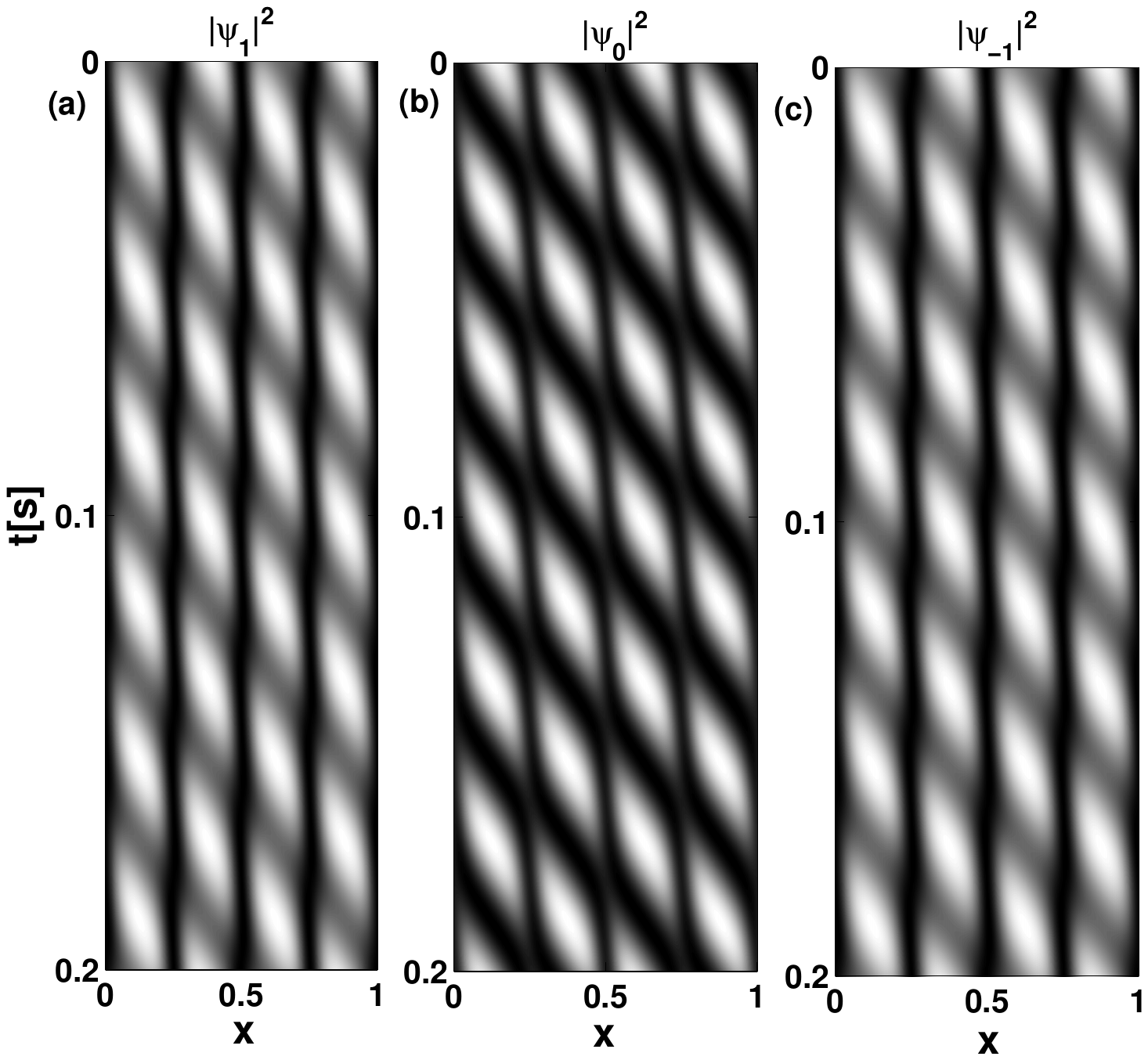}
\caption{The density distribution of the non-stationary solution (\ref{non-D}). The parameters are the
same as in Fig.4.}
\end{center}
\end{figure}

\begin{figure}
\begin{center}
\includegraphics*[width=8.5cm]{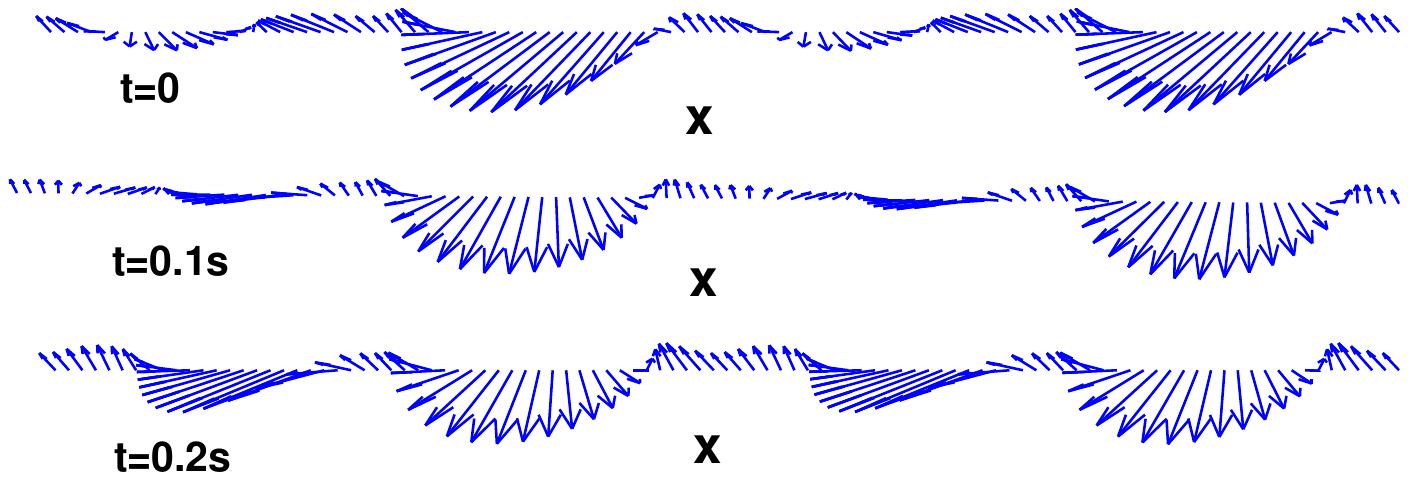}
\caption{Snapshots of the spin-polarization distributions of the non-stationary solutions (\ref{non-D})
at $t=0s, t=0.1s$ and $t=0.2s$, respectively.}
\end{center}
\end{figure}

Fig.(9) and Fig.(11) depict the temporal evolution of the density in
each component from the non-stationary solutions (\ref{non-C}) and
(\ref{non-D}), respectively. They exhibit periodical structures in
both the time and the space. Fig.(10) and Fig.(12) describe the
evolution of the spin-polarizations for the non-stationary solutions
(\ref{non-A}) and (\ref{non-B}) at $t=0, t=0.1s$ and $t=0.2s$,
respectively.

\section{Summary}
In summary, we have analytically presented several types of stationary solutions to the 1D coupled
nonlinear GPEs. Obviously, we can obtain a lot of other stationary solutions with different combination
of the Jacobian elliptic functions. In the meanwhile, we derived exact time-evolution solutions by
making use of the spin-rotational symmetry of the Hamiltonian. We emphasis that the solutions in Sec.II
are stationary. They become non-stationary in Sec.III only after the $SO(3)$ spin-rotation applying to
the equations of motion (\ref{temporal}). The non-zero parameter $\mu_1$ plays an important role in
constructing these non-stationary solutions. Our method is applicable to the $F = 2$ spinor BECs. Works
in this direction are in progress.

This work is supported by the funds from the Ministry of Science and
Technology of China under Grant Nos. 2012CB821403 and by the
National Natural Science Foundation of China under grant No.
10874018.


\begin{thebibliography}{99}
\bibitem{Barrett} M. D. Barrett, J. A. Sauer, and M. S. Chapman, Phys. Rev. Lett. {\bf 87}, 010404 (2001).
\bibitem{Gorlitz} A. G\"{o}rlitz, T. L. Gustavson, A. E. Leanhardt, R. L\"{o}w, A. P. Chikkatur, S. Gupta, S. Inouye, D. E. Pritchard, and W. Ketterle, Phys. Rev. Lett. {\bf 90}, 090401 (2003).
\bibitem{Leanhardt} A. E. Leanhardt, Y. Shin, D. Kielpinski, D. E. Pritchard, and W. Ketterle, Phys. Rev. Lett. {\bf 90}, 140403 (2003).
\bibitem{Stenger} J. Stenger, S. inouye, D. M. Stamper-Kurn, H. J. Miesner, and A. P. Chikkatur, Nature {\bf 396}, 345 (1998).
\bibitem{Miesner} H. J. Miesner, D. M. Stamper-Kurn, J. Stenger, S. Inouye, A. P. Chikkatur, and W. Ketterle, Phys. Rev. Lett. {\bf 82}, 2228 (1999).
\bibitem{Kobayashi} M. Kobayashi, Y. Kawaguchi, M. Nitta, and M. Ueda, Phys. Rev. Lett. {\bf 103}, 115301 (2009).
\bibitem{Ueda} M. Ueda and Y. Kawaguchia, arXiv: 1001.2072
(unpublished).
\bibitem{Kawaguchi} Y. Kawaguchi, H. Saito, and M. Ueda, Phys. Rev. Lett. {\bf 96}, 080405 (2006).
\bibitem{Ho} T.-L. Ho, Phys. Rev. Lett. {\bf 81}, 742 (1998).
\bibitem{Chang} M.-S. Chang, C. D. Hamley, M. D. Barrett, J. A. Sauer, K. M. Fortier, W. Zhang, L. You, and M. S. Chapman, Phys. Rev. Lett. {\bf 92}, 140403 (2004).
\bibitem{Murata} K. Murata, H. Saito, and M. Ueda, Phys. Rev. A {\bf 75}, 013607 (2007).
\bibitem{Imambekov} A. Imambekov, M. Lukin, and E. Demler, Phys. Rev. A {\bf 68}, 063602
(2003).
\bibitem{Gross} E. P. Gross, Phys. Rev. {\bf 106}, 161 (1957).
\bibitem{Ginzburg}V. L. Ginzburg and L.P. Pitaevskii, Sov. phys. JETP {\bf 7}, 858 (1958).
\bibitem{Dalfovo} F. Dalfovo, S. Giorgini, L. P. Pitaevskii, and S. Stringari, Rev. Mod. Phys. {\bf 71}, 463 (1999).
\bibitem{Coen} S. Coen and M. Haelterman, Phys. Rev. Lett. {\bf 87}, 140401 (2001).
\bibitem{N.Z.} N. Z. Petrovi\'{c}, N. B. Aleksic, A. A. Bastami, and M. R. Belic, Phys. Rev. E. {\bf 83}, 036609 (2011).
\bibitem{Li} L. Li, Z. Li, B. A. Malomed, D. Mihalache, and W. M. Liu, Phys. Rev. A {\bf 72},033611 (2005).
\bibitem{Zhang} W. Zhang, \"{O}. E. M\"{u}stecaplio\v{g}lu, and L. You, Phys. Rev. A {\bf 75}, 043601 (2007).
\bibitem{Nistazakis} H. E. Nistazakis, D. J. Frantzeskakis, P. G. Kevrekidis, B. A. Malomed and R. Carretero-Gonz\'{a}lez, Phys. Rev. A {\bf 77},033612(2008).
\bibitem{Gardner} C. S. Gardner, J.  M. Greene, M. D. Kruskal, and R. M.
Miura, Phys. Rev. Lett. {\bf 19}, 1095 (1967).
\bibitem{Sedawy} A. R. Sedawy, Appl. Math. Lett. {\bf 23}, 142 (1973).
\bibitem{Uchiyama} M. Uchiyama, J. Ieda and M. Wadati, J. phys. Soc.
Jpn. {\bf 75}, 064002 (2006).
\bibitem{Ieda} J. Ieda, T. Miyakawa, and M. Wadati, Phys. Rev. Lett. {\bf 93}, 194102 (2004).
\bibitem{Beitia} J. Belmonte-Beitia, V. M. P\'{e}rez-Garc\'{i}a, V. Vekslerchik, and P. J. Torres, Phys. Rev. Lett. {\bf 98}, 064102
(2007).
\bibitem{Theocharis} G. Theocharis, P. Schmelcher, P. G. Kevrekidis, and D. J. Frantzeskakis, Phys. Rev. A {\bf 72}, 033614
(2005).
\bibitem{Avelar} A. T. Avelar, D. Bazeia, and W. B. Cardoso, Phys. Rev. E {\bf 79}, 025602(R)
(2009).
\bibitem{Wang} D.-S. Wang, X.-H. Hu, and W. M. Liu, Phys. Rev. A {\bf 82}, 023612 (2010).
\bibitem{Carr} L. D. Carr, C. W. Clark, and W. P. Reinhardt, Phys. Rev. A. {\bf 62}, 063610 (2000).
\bibitem{Bradley} R. M. Bradley, B. Deconinck, and J. N. Kutz, J. Phys. A: Math. Gen. {\bf 38}, 1901
(2005).
\bibitem{LLi} L. Li, B. A. Malomed, D. Mihalache and W. M. Liu, Phys. Rev. E {\bf 73}, 066610 (2006).
\bibitem{Yan} D. Yan, J. J. Chang, C. Hamner, P. G. Kevrekidis, P. Engels, V. Achilleos, D. J. Frantzeskakis, R. Carretero-Gonzalez, and P. Schmelcher, Phys. Rew. A {\bf 84}, 053630 (2011).
\end{thebibliography}
\end{document}